\documentclass[preprint2]{aastex}
\usepackage{ulem}

\def\ds{\displaystyle}

\shorttitle{Simple analytical approximations for IC process in the black-body radiation field}
\shortauthors{Khangulyan et al.}

\begin{document}

\title{Simple analytical approximations for treatment of inverse Compton scattering of relativistic electrons in the black-body radiation field}

\author{D.~Khangulyan}
\affil{Institute of Space and Astronautical Science/JAXA, \\
3-1-1 Yoshinodai, Chuo-ku, Sagamihara, Kanagawa 252-5210, Japan}
\email{khangul@astro.isas.jaxa.jp}

\author{F.A.~Aharonian}
\affil{Dublin Institute for Advanced Studies, 31 Fitzwilliam Place, Dublin 2, Ireland;
Max-Planck-Institut f\"ur Kernphysik,
Saupfercheckweg 1, D-69117 Heidelberg, Germany}
\email{Felix.Aharonian@mpi-hd.mpg.de}

\author{S.R.~Kelner}
\affil{Max-Planck-Institut f\"ur Kernphysik,
Saupfercheckweg 1, D-69117 Heidelberg, Germany; National Research Nuclear University, Kashira Hwy 31, 115409 Moscow, Russia}
\email{Stanislav.Kelner@mpi-hd.mpg.de}

\begin{abstract}
  The inverse Compton (IC) scattering of relativistic electrons is one
  of the major gamma-ray production mechanisms in different
  environments.  Often the target photons for the IC scattering are
  dominated by black (or grey) body radiation.  In this case, the
  precise treatment of the characteristics of IC radiation requires
  numerical integrations over the Planckian distribution. Formally,
  analytical integrations are also possible but they result in series
  of several special functions; this limits the efficiency of usage of
  these expressions.  The aim of this work is the derivation of
  approximate analytical presentations which would provide adequate
  accuracy for the calculations of the energy spectra of up-scattered
  radiation, the rate of electron energy losses, and the mean energy
  of emitted photons.  Such formulae have been obtained by merging the
  analytical asymptotic limits. The coefficients in these expressions
  are calculated via the least square fitting of the results of
  numerical integrations.  The simple analytical presentations,
  obtained for both the isotropic and anisotropic target radiation
  fields, provide adequate (as good as $1\%$) accuracy for broad
  astrophysical applications.
\end{abstract}


\section{Introduction}
Relativistic electrons can transfer their energy to gamma rays through
the process of inverse Compton (IC) scattering of the ambient low
energy photons. Together with bremsstrahlung, this process represents
a major channel of gamma-ray production by relativistic electrons (and
positrons). At energies below 100~MeV, a non-negligible contribution
to the gamma-ray continuum can be supplied by annihilation of
positrons on flight. In the same energy band, one may expect a
contribution from the synchrotron radiation of electrons. However,
the latter scenario can be effectively realized only in quite unique objects
called extreme accelerators, when the particle acceleration proceeds
at the maximum (theoretically possible) rate. In general, the
synchrotron radiation is released well below the gamma-ray band, and,
in fact, is considered as a dissipative process as long as it concerns
the efficiency of gamma-ray production. Indeed, in environments with
the energy density of the magnetic field significantly exceeding the
energy density of the radiation field, $f= B^2/({8 \pi}w_{\rm
  rad}) \gg 1$, only a small fraction ($f^{-1}$) of the kinetic energy
of electrons is released in high energy gamma rays.  Otherwise, the
radiative cooling of electrons is dominated by the IC scattering,
making the latter an extremely effective gamma-ray production
mechanism, especially at very high energies, when
the radiative cooling {due to} bremsstrahlung is suppressed compared to the
IC scattering\footnote{This follows from the ratio of the cooling
  times due to the bremsstrahlung and IC scattering (in the Thomson
  regime): $t_{\rm br}/t_{\rm IC} \propto E \times
  (N_{\rm gas}/w_{\rm rad})$, where $E$ is the electron's energy, and $N_{\rm gas}$ is the gas density.}.  Because of the large cross-section of
the process and the presence of high density radiation fields, the IC
scattering undoubtedly is the most prolific and universal
gamma-radiation mechanism which contributes from low (MeV) to
ultrahigh (tens of TeV) energies of emission of almost all nonthermal
source populations - supernova remnants (SNR), pulsar wind nebulae
(PWNe), compact binary systems, active galactic nuclei, {\it etc}.

The energy spectrum of the up-scattered photons depends strongly on
the energy of target photon, especially in the Thomson regime,
when the average energy of the up-scattered photon is proportional to the
energy of the target photon. In the Klein-Nishina regime, the most
fraction of the electron energy is transferred to the up-scattered
photon, thus the dependence on the target photon energy gradually
disappears.  Given this nontrivial dependence on the target photon
energy, the accurate calculations of the IC spectrum require good
knowledge of the energy distribution of target photons.  Fortunately,
in many cases the dominant contribution to the IC scattering comes
from photons belonging to black-body (or grey-body) radiation, i.e.  they
are described by the standard Planckian distribution:
\begin{equation}
dN_{\rm ph}\propto{\omega_0^2d\omega_0\over {\rm e}^{\omega_0/T}-1}\,,
\end{equation}
where { $T$ and $\omega_0$  are the photon gas temperature and the photon energy, respectively (hereafter the energies of both photons and electrons, as well as the photon gas temperature,
 are expressed in units of $m_{\rm e}c^2$)}. 

 Planckian distribution is realized  in the case of IC scattering on the 2.7 K cosmic microwave
background radiation (CMBR). Remarkably, in  SNRs and PWNe in our Galaxy, { as well as in
  extragalactic objects like large scale AGN jets and Clusters of
  galaxies,} the IC scattering in the VHE band is strongly dominated
by CMBR.  {The target photons for the IC scattering can be well
  described by the Planckian distribution also in compact systems like
  gamma-ray emitting binaries containing either a pulsar (binary
  pulsars) or a black hole (microquasars). Despite the different
  origins of the nonthermal energy, supported in the first case by a
  rotation-powered pulsar wind, and in the second case by an
  accretion-powered jet, the most likely mechanism of gamma-radiation
  is the IC scattering.  In both type of objects the target photons
  are supplied  by the thermal
radiation of the bright optical star.} Finally, in some cases quite
complex photon distributions can be represented as a superposition of
several grey-body components.

Since the Planckian distribution of photons is characterized by a rapid
decrease of the density both at low and at high energies, numerical
integrations of the Compton cross section over the Planckian photon field
{generally} do not impose  computational difficulties, but simply require
additional computational time. Often, to shorten the calculations,
different approximations are used. The most common approach is the
$\delta$-functional approximation for the narrow Planckian spectrum.
This approximation can correctly  describe the lower energy part of the spectrum,
but is not applicable for the precise computations of the entire gamma-ray spectra (see
Section~\ref{sec:comp}). Therefore, some other approximations for calculations of IC spectra have been
recently suggested in the
literature \citep[see][]{petruk09,zdz13}. These approaches provide
a better description for the IC spectra and are characterized by a
higher precision than the $\delta$-functional approximation does,
although they are not free of certain limitations (see discussion in
Section~\ref{sec:comp}).

In this paper we propose new, very simple analytical presentations
obtained for both isotropic and mono-directional angular distributions of
target radiation field assuming that its energy spectrum is precisely
described by the Planckian distribution. We provide analytical
formulae for the energy spectra of upscattered radiation, as well as
for the interaction and the energy-loss rates of electrons.  The
``threshold'' for the accuracy of these formulae in all cases has been
set at the level of $1\%$.

\section{Approximate description of the IC process}\label{sec:cross_section}
The interaction of electrons with photons is described with the
standard means of quantum electrodynamics. In the astrophysical
context, the general expressions for the Compton cross-section can be
significantly simplified using the fact that the energy of the target
photons is typically small, $\omega_0\ll 1$, and the electrons are
relativistic, $E\gg1$ { (this condition, in particular, implies that the up-scattered photon moves in the direction of the initial velocity of the electron)}. Under these
circumstances, for the target photons with a fixed direction, the
scattering rate by an electron moving with a velocity which makes an
angle $\theta$ with the photon's direction has the following simple
form \citep{aa81}:
\begin{equation}\label{eq:aa_ani}
\begin{array}{rcl}
 {d\nu_{\rm ani}\over d \omega\,dN_{\rm ph}\,dt}&=& c\left(1-\cos\theta\right){d\sigma\over d\omega}\\
&=& {4\pi cr_0^2\left(1-\cos\theta\right)\over b_\theta E}\times\\
\multicolumn{3}{c}{ \left[{1+\frac{z^2}{2(1-z)}-\frac{2z}{b_\theta(1-z)}+\frac{2z^2}{b_\theta^2(1-z)^2}}\right]\,,}
\end{array}
\end{equation}
where  $E$, $\omega_0=b_\theta/\left[2E(1-\cos\theta)\right]$ and  $\omega=zE$ are energies of electron, soft photon,  and up-scattered photon, respectively; $r_0=e^2/(m_{\rm e}c^2)$ is the electron classical radius; and $N_{\rm ph}$ is {number of target photons per unit of volume}.  If the target photon field is isotropic, the above expression should be averaged over the interaction angle \citep{aa81}:
\begin{equation}\label{eq:aa_iso}
\begin{array}{rcl}
{d\nu_{\rm iso}\over d \omega\,dN_{\rm ph}\,dt}&=& c\int\left(1-\cos\theta\right){d\sigma\over d\omega}\,{d\Omega\over 4\pi}\\
&=&{8\pi cr_0^2\over bE}\times\\
\multicolumn{3}{c}{ \left[1+\frac{z^2}{2(1-z)}+\frac{z}{b(1-z)}-\frac{2z^2}{b^2(1-z)^2}-\right.}\\
\multicolumn{3}{c}{\left.\frac{z^3}{2b(1-z)^2}-\frac{2z}{b(1-z)}\,\log{b(1-z)\over z}\right]\,,}
\end{array}
\end{equation}
where $b=4\omega_0 E$ is the Klein-Nishina parameter and the notation $\log$ is used for logarithm to the base $e$, i.e. ``natural logarithm''. { Note that Equation (\ref{eq:aa_iso}) has  been originally derived by \citet{jones68} in a straight way, without using the intermediate  angle-dependent rate given by 
 Equation~(\ref{eq:aa_ani}) \citep[for a review see][]{bg70}.}

In the case of black-body target photons, the scattering rates given by Equations~(\ref{eq:aa_ani}) and (\ref{eq:aa_iso}) should be integrated over the Planckian distribution of target photons:
\begin{equation}\label{eq:planck}
dN_{\rm ph}={m_{\rm e}^3c^3\kappa\over\pi^2\hbar^3}{\omega_0^2\,d\omega_0\over {\rm e}^{\omega_0/T}-1}\,,%,
\end{equation}
where $\kappa$ is the dilution factor in the case of grey-body radiation.
The lower integration limit, $\omega_0\geq\epsilon_{\rm ani/iso}$,  is determined by the kinematic conditions (i.e., conditions imposed by the conservation of 4-momentum) as 
\begin{equation}
\epsilon_{\rm ani}=\frac\omega{2E}\frac1{(E-\omega)}\frac1{(1-\cos\theta)}={z\over1-z}{T\over t_\theta}
\end{equation}
and 
\begin{equation}
\epsilon_{\rm iso}=\frac\omega{4E}\frac1{(E-\omega)}={z\over1-z}{T\over t}
\end{equation}
for the cases of mono-directional and isotropic photon distributions, respectively. Here, the following notations are used: $t_\theta=2ET(1-\cos\theta)$ and $t=4ET$.

Formally, for large values of $\omega_{0}\rightarrow+\infty$, Equations~(\ref{eq:aa_ani}) and (\ref{eq:aa_iso}) are not applicable, since the basic assumption, $\omega_0\ll\omega$, used for the derivation of these expressions fails.  However, assuming a non-relativistic photon temperature $T\ll1$, one can safely extend the integration upper limit to $+\infty$. Therefore, the interaction rate with black-body distribution of target photons is 
\begin{equation}\label{eq:bb_ic}
{dN_{\rm ani/iso}\over d \omega\,dt}= {T^3m_{\rm e}^3c^3\kappa\over\pi^2\hbar^3}\int\limits_{\epsilon_{\rm ani/iso}/T}^\infty{d\nu_{\rm ani/iso}\over d \omega\,dN_{\rm ph}\,dt}{x^2dx\over{\rm e}^{x}-1}\,.
\end{equation}

The substitution of Equations~(\ref{eq:aa_ani}) and (\ref{eq:aa_iso}) into Equation~(\ref{eq:bb_ic}) leads to an expression that can be presented in the form which contains the following functions introduced by \citet{zdz13} ({Equations~(15)~and~(29)} in their paper):
\begin{equation}\label{eq:f_pm}
f_{i}=\int\limits_{x_0}^\infty {x^{i}\,dx\over {\rm e}^x-1}\,,
\end{equation}
for $i=-1$, $0$ and $+1$; and 
\begin{equation}\label{eq:f_ln}
f_{\rm ln}=\int\limits_{x_0}^\infty {\log(x)\, dx\over {\rm e}^x-1}\,.
\end{equation}
While the integral $f_0$ is expressed through elementary functions: 
\begin{equation}
f_0=-\log\left(1-{\rm e}^{-x_0}\right)\,,
\end{equation}
and $f_{+1}$ allows a representation with the dilogarithm function \citep[see Equation(7) of][]{zdz13}, the two other
functions ($f_{-1}$ and $f_{\rm ln}$), as shown by \citet{zdz13}, can be expressed through series
\citep[Equation~(14,~28) of][]{zdz13}. {However, each of these terms contains special functions, which makes the
  usage of these expressions rather inconvenient.

  We use a different approach to obtain approximate formulae for the IC cross-section. First, we present the cross-section in a form
  containing strictly positive terms.  This dramatically reduces the risk of large mistakes because of the summation
  of rounding errors. Furthermore, we approximate these terms by simple analytical expressions. The derivations of
  these expressions are based on analytical computations of the asymptotic limits and the introduction of correction
  functions for the intermediate range of energies by invoking a least square fit.  This allows us to present
  Equation~(\ref{eq:bb_ic}) in a form that contains only elementary functions.

}

For the case of a mono-directional photon field, Equation~(\ref{eq:bb_ic}) can be expressed as:
\begin{equation}\label{eq:cross_ani}
\begin{array}{l}
\ds \quad\quad{dN_{\rm ani}\over d \omega\,dt}=\\[10pt]
\ds{2r_o^2m_{\rm e}^3c^4\kappa T^2\over \pi\hbar^3E^2}\times\bigg[{z^2\over2(1-z)}F_{1}\left(x_0\right)+F_2\left(x_0\right)\bigg]\,,
\end{array}
\end{equation}
where  $x_0={z\over(1-z)t_\theta}$, and the positive functions $F_{1}$ and $F_{2}$ are determined as:
\begin{eqnarray}
F_1(x_0)&=&f_{+1}\left(x_0\right)\,,\label{eq:f_1}\\
F_2(x_0)&=&f_{+1}\left(x_0\right)+\nonumber\\
&&2x_0^2f_{-1}\left(x_0\right)-2x_0f_{0}\left(x_0\right)\,.\label{eq:f_2}
\end{eqnarray}

For the case of an isotropic photon field,  Equation~(\ref{eq:bb_ic}) obtains the following form:
\begin{equation}\label{eq:cross_iso}
\begin{array}{l}
\ds\quad\quad {dN_{\rm iso}\over d \omega\,dt}=\\[10pt] \ds {2r_o^2m_{\rm e}^3c^4\kappa T^2\over \pi\hbar^3E^2}\times
\bigg[{z^2\over2(1-z)}F_3(x_0)+F_4(x_0)\bigg]\,,
\end{array}
\end{equation}
where  $x_0={z\over(1-z)t}$. The positive functions $F_{3}$ and $F_{4}$ are expressed via $f_{-1}$, $f_{0}$, $f_{+1}$ and $f_{\rm ln}$:
\begin{equation}\label{eq:f_3}
F_3(x_0)=f_{+1}(x_0)-x_0f_0(x_0)\,;
\end{equation}
and
\begin{equation}\label{eq:f_4}
\begin{array}{rcl}
F_4(x_0)&=&\bigg(f_{+1}(x_0)-x_0f_0(x_0)\bigg)+\\
&\multicolumn{2}{c}{2x_0\bigg(\log(x_0)f_0(x_0)-f_{\rm ln}(x_0)\bigg)+}\\
&&2x_0\bigg(f_0(x_0)-x_0f_{-1}(x_0)\bigg)\,.
\end{array}
\end{equation}

Thus, the IC radiation spectra 
can be presented in simple analytical forms of Equations~(\ref{eq:cross_ani})~and~(\ref{eq:cross_iso}) as  functions  of the variable  $z$
through the term  $z^2/(2(1-z))$ and four functions $F_1$, $F_2$, $F_3$ and $F_4$. All these functions depend only on the parameter $x_0$, which is equal to $z/((1-z)t_\theta)$ and  $z/((1-z)t)$ in the case of mono-directional and isotropic photon fields, respectively.

\begin{figure}
\plotone{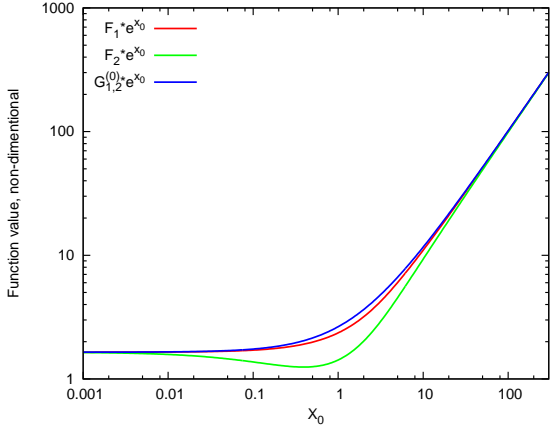}
  \caption{ Functions $F_{1}$, $F_{2}$ are shown together with the zeroth-order approximation $G_{1}^{(0)}$(the function $G_{2}^{(0)}$ is equal to $G_{1}^{(0)}$). The function $G_1^{(0)}$ reproduces asymptotic behavior of functions $F_{1,2}$, but differs by a factor $\lesssim2$ for $x_0\sim1$.}
  \label{fig:f_12}
\end{figure}

The functions $F_1$ and $F_2$ have the same asymptotic limits:
\begin{equation}
\begin{array}{rclc}
\ds F_{1,2}&=&\ds \left\{{\frac{\pi^2}6, \atop x_0{\rm e}^{-x_0},}\right.& \ds { x_0\ll 1
 \atop  x_0\gg 1}\,.
\end{array}
\end{equation}
A simple function with the similar asymptotic behavior can be used as a zeroth-order approximation for the functions $F_1$ and $F_2$ (see Figure~\ref{fig:f_12}):
\begin{equation}
G_{1,2}^{(0)}=\left(\frac{\pi^2}6+x_0\right){\rm e}^{-x_0}\,.
\end{equation}
A numerical comparison of these functions shows that $G_{1}^{(0)}$  provides\footnote{The functions $G_{1}^{(0)}$ and $G_{2}^{(0)}$ are identical.} accuracy of $10\%$ and $50\%$ for $F_1$ and $F_2$, respectively. To improve the accuracy, we introduce correction functions:
\begin{equation}\label{eq:app_pm}
G_{1,2}=G_{1,2}^{(0)}\left(x_0\right)\times g_{1,2}\left(x_0\right)\,,
\end{equation}
We present the correction factors $g_1$ and $g_2$ as functions of the variable $x_0$ with four free parameters:
\begin{equation}\label{eq:correction}
g_{i}\left(x_0\right)=\left[1+\frac{a_{i}x_0^{\alpha_{i}}}{1+b_{i}x_0^{\beta_{i}}}\right]^{-1}\,.
\end{equation}
The parameters $\alpha_{i}>0,a_{i},\beta_{i}>\alpha_{i},b_{i}>0$ were used for the least-square fitting of functions $F_{1,2}$. Obviously, this is not a unique representation for the approximation function, but since $g\rightarrow1$ for $x\ll1$ and $x\gg1$, the considered function family should preserve the asymptotic behavior of the zeroth-order fit $G_{1,2}^{(0)}$  and provide enough freedom for fitting. We therefore use Equation~(\ref{eq:correction}) as the correction function for fitting procedures used in this paper.

\begin{figure}
\plotone{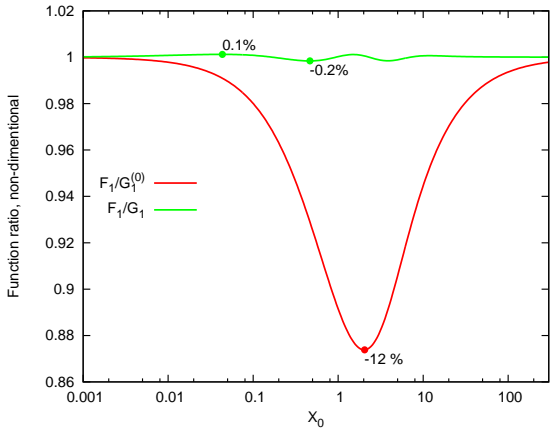}
  \caption{The ratios of function $F_1$ to approximations $G_1^{(0)}$ and $G_1$.}
  \label{fig:f_1ratio}
\end{figure}

\begin{figure}
\plotone{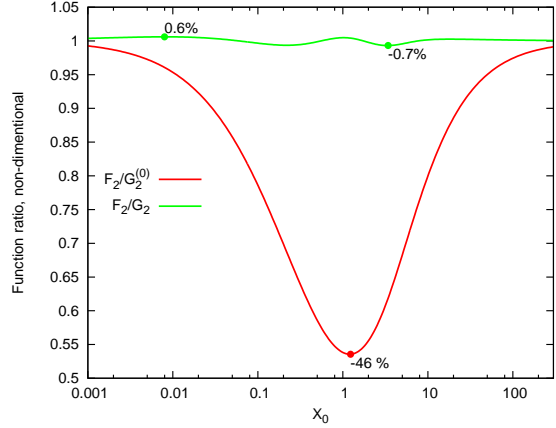}
  \caption{The ratios of function $F_2$ to approximations $G_2^{(0)}$ and $G_2$.}
  \label{fig:f_2ratio}
\end{figure}

The numerical least square fitting gives the following sets of parameters
\begin{equation}\label{eq:par1}
\begin{array}{rclrcl}
\alpha_{1}&=&0.857,& a_{1}&=&0.153,\\ \beta_{1}&=&1.84,& b_{1}&=&0.254\,;
\end{array}
\end{equation}
and
\begin{equation}\label{eq:par2}
\begin{array}{rclrcl}
\alpha_{2}&=&0.691,& a_{2}&=&1.33,\\ \beta_{2}&=& 1.668,& b_{2}&=&0.534\,,
\end{array}
\end{equation}
which provide a precision of better than $1\%$ for the entire range of $x_0$, as shown in Figures~\ref{fig:f_1ratio}~and~\ref{fig:f_2ratio}.

\begin{figure}
\plotone{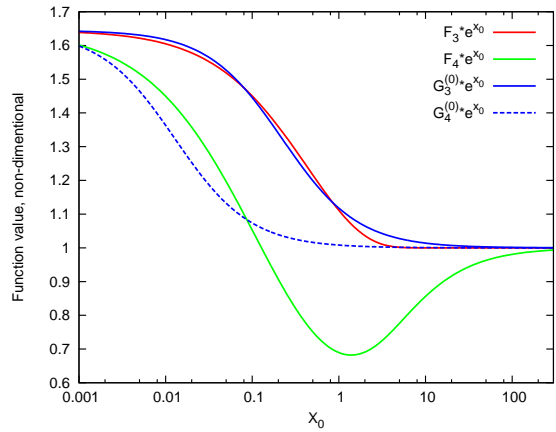}
  \caption{ Functions $F_{3}$, $F_{4}$ are shown together with the zeroth-order approximations $G_{3}^{(0)}$, computed for $c_3=2.73$, and $G_{4}^{(0)}$, computed for $c_4=47.1$. % The function $G_3^{(0)}$ reproduces asymptotic behavior of functions $F_{3,4}$, but differs from $F_4$ by a factor $\lesssim2$ for $x_0\sim1$.
  }
  \label{fig:f_34}
\end{figure}

A similar approach can be used for the approximation of the angle averaged  IC spectra determined by Equation~(\ref{eq:cross_iso}). The functions $F_{3}$ and $F_{4}$ have the same asymptotic: 
\begin{equation}
\begin{array}{rclc}
\ds F_{3,4}&=&\ds \left\{{\frac{\pi^2}6, \atop {\rm e}^{-x_0},}\right.&\ds { x_0\ll 1
 \atop  x_0\gg 1}\,,
\end{array}
\end{equation}
which suggests the following family of approximation functions:
\begin{equation}\label{eq:app_34}
G_{\rm 3,4}=G_{3,4}^{(0)}(x_0)\times g_{3,4}(x_0)\,,
\end{equation}
where
\begin{equation}\label{eq:app_30}
G_{\rm 3,4}^{(0)}(x_0)=\frac{\pi^2}6\frac{1+c_{3,4}\,x_0}{1+\frac{\pi^2c_{3,4}}6x_0}{\rm e}^{-x_0}\,.
\end{equation}
The functions $G_{\rm 3,4}^{(0)}$ have a similar asymptotic behavior as the functions $F_{3,4}$; it is demonstrated in Figure~\ref{fig:f_34}. Here $c_{3,4}>0$ are parameters which do not change the asymptotic behavior of $G_{\rm 3,4}^{(0)}$. Therefore, they can be optimized for a better description of functions $F_{3,4}$. In particular, for the value of $c_3=2.73$ the function $G_{\rm 3}^{(0)}$ provides a $3\%$ accuracy for the function $F_{\rm 3}$, as shown in Figure~\ref{fig:f_34}. 

\begin{figure}
\plotone{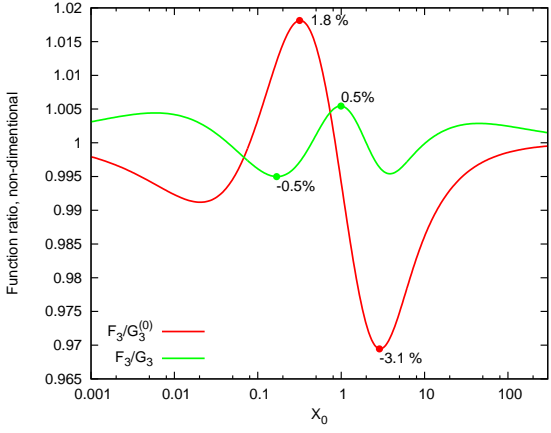}
  \caption{The ratios of function $F_3$ to approximations $G_3^{(0)}$ (taken for $c_3=2.73$) and $G_3$. The used fitting parameters are shown by Equation~(\ref{eq:par3}).}
  \label{fig:f_3ratio}
\end{figure}

\begin{figure}
\plotone{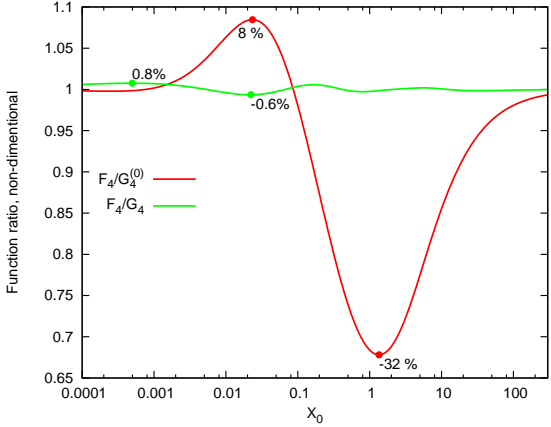}
  \caption{The ratios of the function $F_4$ to approximations $G_4^{(0)}$ (taken for $c_4=47.1$) and $G_4$.  We show this plot for increased range of $x_0$, as compared to Figures~\ref{fig:f_12}~--~\ref{fig:f_3ratio}, to include the region where the ratio of $F_4$ to $G_4$ achieves its maximum.}
  \label{fig:f_4ratio}
\end{figure}

As can be seen in Figure~\ref{fig:f_34}, the function $F_4\times {\rm e}^{x_0}$ features a $\sim30\%$ dip at $x_0\sim1$, which cannot be reproduced by the function $G_4^{(0)}$. Therefore the parameter $c_4$ alone cannot provide an approximation with precision better than a $30\%$  for the function $F_4$ (such accuracy can be achieved for $c_4\simeq50$). However, Equation~(\ref{eq:app_34}) provides a five-parameter ($a_{3,4}$, $\alpha_{3,4}$, $b_{3,4}$, $\beta_{3,4}$ and $c_{3,4}$) function family that can be used for fitting functions $F_{3,4}$. The numerical least-square fits resulted in the following sets of parameters
\begin{equation}\label{eq:par3}
\begin{array}{rclrcl}
\alpha_{3}&=&0.606,&a_{3}&=&0.443,\\
\beta_{3}&=&1.481,& b_{3}&=&0.54,\\
&&&c_{3}&=&0.319\,;
\end{array}
\end{equation}
and
\begin{equation}\label{eq:par4}
\begin{array}{rclrcl}
\alpha_{4}&=&0.461,& a_{4}&=&0.726,\\
\beta_{4}&=& 1.457,&b_{4}&=&0.382,\\
&&& c_4&=&6.62\,,
\end{array}
\end{equation}
which gave a $<1\%$ precision  for the entire range of $x_0$, as shown in Figures~\ref{fig:f_3ratio}~and~\ref{fig:f_4ratio}.

The parameterizations  for the functions $F_{1}$, $F_{2}$, $F_{3}$, $F_{4}$ given by Equations~(\ref{eq:app_pm}) and (\ref{eq:app_34}), with corresponding parameters from Equations~(\ref{eq:par1}--\ref{eq:par2}) and (\ref{eq:par3}--\ref{eq:par4}),  allow us to describe the IC spectra, Equations~(\ref{eq:cross_ani})~and~(\ref{eq:cross_iso}),  with a  precision better than $1\%$. Note that this value corresponds to the maximum deviation of the approximate formulae from the precise value; in the case of a broad distribution of electrons, the accuracies obviously will be significantly better.

The obtained above approximations describe the IC spectra for two different,  isotropic and mono-directional angular distributions of target photons. The former scenario  with an involvement of CMBR is often realized in different objects  like  SNRs, PWNe, and Clusters of Galaxies. However, in many other cases the background photon field can be approximated as a grey-body (or a superposition of a several grey-body components) emission. In this case the energy, $\omega_*$, at which the {spectral energy distribution (i.e., $\nu F_\nu$)} of target photons achieves the maximum, allows to define the temperature of the grey-body emission:
\begin{equation}
T=0.255\,\omega_*\,.
\end{equation}
The energy density of the target photon field, $w_*$, allows then to obtain the corresponding dilution coefficient:
\begin{equation}
\kappa=360{\hbar^3w_*\over m_{\rm e}^3c^5}\,\,\left({\omega_*\over m_{\rm e}c^2}\right)^{-4}\,.
\end{equation}

The approximation of a mono-directional photon field is applicable when  the source of the target photons is compact, namely  in the  IC production region the value of $(1-\cos\theta)$ should not vary significantly for the photons coming from different regions of the source. Then in the IC production site, the target photon field is typically diluted by a factor
\begin{equation}\label{eq:pl_dilution}
\kappa={\Delta\Omega\over 4\pi}\,,
\end{equation}
where $\Delta\Omega\ll1$ is the solid angle of the target photons' source, as seen from the IC production region. In case the source of target photons is a star, Equation~(\ref{eq:pl_dilution}) can be expressed through the radius,  $R_*$, of the star and distance, $R$, between the IC emitter and the star {(the condition $\Delta\Omega\ll1$ is realized if $R\gg R_*$)}:
\begin{equation}
\kappa=\left({R_*\over 2R}\right)^2\,.
\end{equation}

\section{The rate of IC losses}\label{sec:losses}
IC energy losses of a particle in a Planckian photon field can be expressed as
\begin{equation}\label{eq:bb_losses}
\begin{array}{l}
\quad\quad{\dot{E}_{\rm ani/iso}}= {m_{\rm e}^3c^3\kappa E^2\over\pi^2\hbar^3}\times\\\int\limits_0^\infty\,d\omega_0\,{\omega_0^2\over{\rm e}^{\omega_0/T}-1} \int\limits_0^{z_{\rm max}}\, dz\, z\,{d\nu_{\rm ani/iso}\over d \omega\,dN_{\rm ph}\,dt}\,,
\end{array}
\end{equation}
where $z_{\rm max}$ is $b\over 1+b$ or $b_\theta\over 1+b_\theta$ for isotropic and mono-directional photon fields, respectively. Formally, the integration lower limit is not equal to $0$, {however} the contribution to the integral from the region of small $z$ is negligible. Therefore we formally set the lower limit to $0$.

\begin{figure}
\plotone{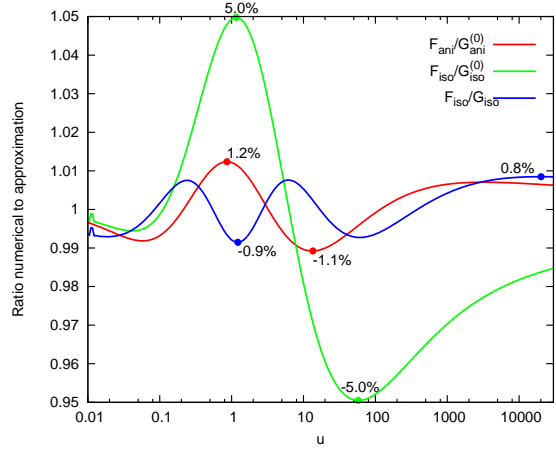}
  \caption{The ratio of the function $F_{\rm ani}$ to $G_{\rm ani}^{(0)}$ (for $c_{\rm ani}=6.13$) and  $F_{\rm iso}$ to $G_{\rm iso}^{(0)}$ (for $c_{\rm iso}=4.62$) and $G_{\rm iso}^{(0)}\times g(u)$.}
  \label{fig:losses}
\end{figure}

The structure of Equation~(\ref{eq:bb_losses}) allows us to determine the dependence of the energy-loss rate on the electron energy and photon field temperature. Namely, if the IC scattering proceeds with a fixed interaction angle, one obtains 
\begin{equation}\label{eq:losses_ani}
  \begin{array}{ll}
    \dot{E}_{\rm ani}&= {r_0^2m_{\rm e}^3c^4\kappa \over2\pi\hbar^3\left({E\left(1-\cos\theta\right)}\right)^{2}}\int\limits_0^\infty\,db_\theta\,{b_\theta\over{\rm e}^{b_\theta/t_\theta}-1}\times\\
\multicolumn{2}{c}{\int\limits_0^{b_\theta\over(1+b_\theta)}\,dz\,z\left[{1+\frac{z^2}{2(1-z)}-\frac{2z}{b_\theta(1-z)}+\frac{2z^2}{b_\theta^2(1-z)^2}}\right]}\\[10pt]
&\ds ={2r_0^2m_{\rm e}^3c^4\kappa T^2\over\pi\hbar^3}F_{\rm ani}\left(t_\theta\right)\,.
  \end{array}
\end{equation}
Based on the asymptotic behavior of the function $F_{\rm ani}$
\begin{equation}\label{eq:ani_losse_asym}
\begin{array}{lc}
\ds F_{\rm ani}(u)=\left\{{\frac{\pi^4}{45}u^2=2.16u^2, \atop \frac{\pi^2}{12}\,\log(u)=0.822\log(u),}\right.&\ds { u\ll 1
 \atop  u\gg 1}\,,
\end{array}
\end{equation}
we suggest the following approximate presentation for the function  $F_{\rm ani}$
\begin{equation}\label{eq:losses_ani_zero}
G^{(0)}_{\rm ani}(u)={c_{\rm ani}\,u\log(1+2.16u/c_{\rm ani})\over1+c_{\rm ani}\,u/0.822}\,.
\end{equation}
Least square fitting for the parameter $c_{\rm ani}$ results in $c_{\rm ani}=6.13$, for which Equation~(\ref{eq:losses_ani_zero}) provides accuracy of an order of $1\%$ (see Figure~\ref{fig:losses}).

Similarly, if IC cooling proceeds with isotropized scattering angles, the energy loss rate is 
\begin{equation}\label{eq:losses_iso}
  \begin{array}{rcl}
    \dot{E}_{\rm iso}&=&{r_0^2m_{\rm e}^3c^4\kappa \over8\pi\hbar^3E^2}\int\limits_0^\infty\,db\,{b\over{\rm e}^{b/t}-1}\times\\
\multicolumn{3}{c}{\int\limits_0^{b\over(1+b)}\,dz\,z\left[1+\frac{z^2}{2(1-z)}+\frac{z}{b(1-z)}-\right.}\\
\multicolumn{3}{c}{\left.\frac{2z^2}{b^2(1-z)^2}-\frac{z^3}{2b(1-z)^2}-\frac{2z}{b(1-z)}\,\log{b(1-z)\over z}\right]}\\[10pt]
&\ds =&\ds{2r_0^2m_{\rm e}^3c^4\kappa T^2\over\pi\hbar^3}F_{\rm iso}(t)\,.
  \end{array}
\end{equation}
The asymptotic behavior of function $F_{\rm iso}$ is similar to Equation~(\ref{eq:ani_losse_asym}):
\begin{equation}\label{eq:ani_losse_iso_asym}
\begin{array}{lc}
\ds F_{\rm iso}(u)=\left\{{\frac{\pi^4}{135}u^2=0.722u^2, \atop \frac{\pi^2}{12}\,\log(u)=0.822\log(u),}\right.&\ds{ u\ll 1
 \atop  u\gg 1}\,,
\end{array}
\end{equation}
Therefore, we can use a function similar to  Equation~(\ref{eq:losses_ani_zero}):
\begin{equation}\label{eq:losses_iso_zero}
G^{(0)}_{\rm iso}(u)={c_{\rm iso}u\log(1+0.722u/c_{\rm iso})\over1+c_{\rm iso}u/0.822}\,.
\end{equation}
Least square fitting for this function renders a value of $c_{\rm iso}=4.62$, for which Equation~(\ref{eq:losses_iso_zero}) provides a $\sim5\%$ precision (see Figure~\ref{fig:losses}). This approximation can be improved by using the correction function defined by  Equation~(\ref{eq:correction}). In particular, a set of the parameters 
\begin{equation}\label{eq:par_iso}
\begin{array}{rclrcl}
\alpha_{\rm iso}&=&0.682,& a_{\rm iso}&=&-0.362,\\
\beta_{\rm iso}&=&1.281,& b_{\rm iso}&=&0.826,\\
&&& c_{\rm iso}&=&5.68
\end{array}
\end{equation}
provide a $\sim1\%$ precision (see Figure~\ref{fig:losses}). A similar expression (although less optimized, with an accuracy of $3\%$) was originally presented in \citet{bk09}.

The obtained Equation~(\ref{eq:losses_ani_zero})
for $F_{\rm ani}$ and
Equations~(\ref{eq:losses_iso_zero}~--~\ref{eq:par_iso}) for $F_{\rm iso}$
allow a precise description of the energy losses with formulae
Equations~(\ref{eq:losses_ani})~and~(\ref{eq:losses_iso}), which correspond to
the case of electron-photon interactions at a specific angle and angle
averaged, respectively. It is important to note that the latter
can be realized also in the cases when the target photons are mono-directional in the
source frame. For example, if particles are isotropized by the source
magnetic field, the production of the IC emission towards the observer
proceeds at a specific interaction angle, since relativistic particles
emit within a narrow cone towards the direction of their motion. However,
if IC cooling time exceeds particle isotropization timescale, each
particle can interact with photons at an arbitrary angle,
therefore particle losses are effectively determined by the interaction with an isotropic
photon field, i.e., by Equation~(\ref{eq:losses_iso}).

The particle cooling described by
Equation~(\ref{eq:losses_ani}) can be realized, for example, in the so-called Compton-drag scenarios, i.e. when  the temperature of the emitting
particles is very small and the bulk motion component is dominant. In
particular, this may be relevant to the  pulsar-wind zone for pulsars
located in systems with bright stars \cite[see,
e.g.,][]{kha07,kab11}. Also a similar situation can arise if IC
cooling time is shorter than  isotropization time-scale.

Often it is convenient to characterize the energy losses through the cooling-time:
\begin{equation}\label{eq:cooling}
t_{\rm ic}={E\over\dot{E}}={\pi\hbar^3\over2r_0^2m_{\rm e}^3c^4\kappa T^2}{E\over F_{\rm ani/iso}\left(t_{\theta/\cdot}\right)}\,.
\end{equation}
In the Thomson limit, for the case of an isotropic photon field, this expression gives:
\begin{equation}
t_{\rm ic}=\left(\frac{4cE}3{\pi^2\kappa T^4m_{\rm e}^3c^3\over 15\hbar^3}\frac{8\pi r_0^2}3\right)^{-1}\,,
\end{equation}
where the middle term corresponds to the ratio of energy density of the black-body photon distribution to $m_{\rm e}c^2$.

It was suggested in \citet{hess06} and
further generalized by \cite{bk09} that, {in the case of an isotropic photon field,} IC cooling time in the
Klein-Nishina limit can be described by a simple  function:
\begin{equation}\label{eq:cooling_07}
t_{\rm ic}\approx5\times10^{-17}T^{-2.3}E^{0.7}\rm s\,,
\end{equation}
where we transformed the numerical coefficient from Equation~(13) in
\cite{bk09} to the units used in this paper, and adopted a dilution
coefficient to be $1$. The comparison of Equations~(\ref{eq:cooling})~and~(\ref{eq:cooling_07}) shows
that the latter implies
that function $F_{\rm iso}$ has been approximated as $F_{\rm
  iso}\simeq0.4t^{0.3}$, which provides an accuracy of $<30\%$ for
$5<t<10^3$.

\section{Interaction rate}\label{sec:rate}
Another important characteristic of Compton scattering {are} the {interaction rates:}
\begin{equation}\label{eq:bb_rate_ani}
\begin{array}{l}
\quad\quad{\dot{N}_{\rm ani}}= \\
{m_{\rm e}^3c^4\kappa \over\pi^2\hbar^3}\int\limits_0^\infty\,d\omega_0\,{\omega_0^2\over{\rm e}^{\omega_0/T}-1} \left(1-\cos\theta\right)\sigma\left(b_\theta\right)\,,
\end{array}
\end{equation}
and
\begin{equation}\label{eq:bb_rate_iso}
\begin{array}{l}
\quad\quad{\dot{N}_{\rm iso}}= \\{m_{\rm e}^3c^4\kappa \over\pi^2\hbar^3}\int\limits_0^\infty\,d\omega_0\,{\omega_0^2\over{\rm e}^{\omega_0/T}-1} \int\limits_{-1}^{1}\,{d\cos\theta\over2}\,\left(1-\cos\theta\right)\sigma\left(b_\theta\right)\,,
\end{array}
\end{equation}
{for the scattering at a fixed interaction angle,} and for the angle averaged interactions, respectively. 
Here $\sigma$ is the Lorentz-invariant cross section for Compton scattering \citep[see e.g.][]{landau4}:
\begin{equation}\label{eq:ic_cros}
\begin{array}{l}
\quad\quad\sigma(x)={2\pi r_0^2\over x}\times\\
\left[\left(1-\frac4x-\frac8{x^2}\right)\log(1+x)+\frac12+\frac8x-\frac1{2(1+x)^2}\right]\,.
\end{array}
\end{equation}

\begin{figure}
\plotone{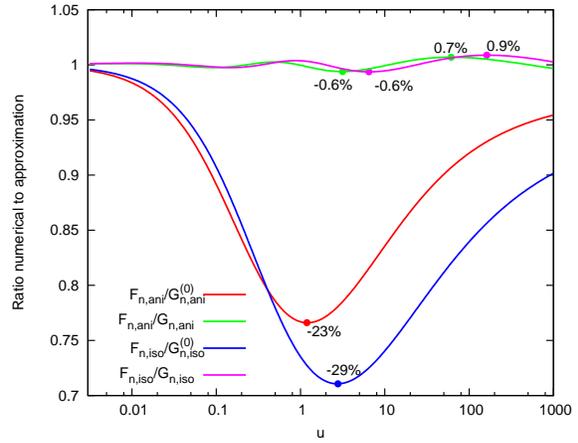}
  \caption{The ratios of the function $F_{\rm n,ani}$ to $G_{\rm n,ani}^{(0)}$ and to $G_{\rm n,ani}^{(0)}\times g(u)$; and the ratios of the function  $F_{\rm n,iso}$ to $G_{\rm n,iso}^{(0)}$  and to $G_{\rm n,iso}^{(0)}\times g(u)$.}
  \label{fig:rate}
\end{figure}

For the case of fixed scattering angle, the interaction rate can be expanded as
\begin{equation}\label{eq:bb_rate_ani_exp}
\begin{array}{rcl}
{\dot{N}_{\rm ani}}&=& {r_0^2m_{\rm e}^3c^4\kappa \over4\pi\hbar^3E^3\left(1-cos\theta\right)^2}\times\\
\multicolumn{3}{c}{\int\limits_0^\infty\,db_\theta\,{b_\theta\over{\rm e}^{b_\theta/t_\theta}-1}\left[\left(1-\frac4{b_\theta}-\frac8{{b_\theta}^2}\right)\log(1+{b_\theta})+\right.}\\
\multicolumn{3}{c}{\left.\frac12+\frac8{b_\theta}-\frac1{2(1+{b_\theta})^2}\right]}\\[10pt]
&\ds=&\ds{2r_0^2m_{\rm e}^3c^4\kappa T^2\over\pi\hbar^3E}F_{\rm n,ani}\left(t_\theta\right)\,.
\end{array}
\end{equation}
The integral term in Equation~(\ref{eq:bb_rate_ani_exp}) has the following asymptotic behavior\footnote{Here $\zeta$ {denotes the Riemann Zeta-function.}}:
\begin{equation}\label{eq:nani_asym}
\begin{array}{lc}
\ds F_{\rm n,ani}(u)=\left\{{\frac43\zeta(3)\,u=1.6\,u, \atop \frac{\pi^2}{12}\,\log(u)=0.822\log(u),}\right.&\ds{ u\ll 1
 \atop  u\gg 1}\,,
\end{array}
\end{equation}
therefore in the zeroth-order approximations this function can be presented in the form:
\begin{equation}\label{eq:G0_nani}
G_{\rm n,ani}^{(0)}(u)=0.822\log\left(1+1.95u\right)\,.
\end{equation}
Like in the previous cases, one can improve this approximate formula by the correction function given in Equation~(\ref{eq:correction}). In Figure~\ref{fig:rate}, we show $G_{\rm n,ani}^{(0)}(u)\times g(u)$ for the following parameter values:
\begin{equation}\label{eq:par_ani_rate}
\begin{array}{rclrcl}
\alpha_{\rm n,ani}&=&0.885,& a_{\rm n,ani}&=&1.05,\\
\beta_{\rm n,ani}&=&1.213,& b_{\rm n,ani}&=&2.46\,
\end{array}
\end{equation}
It can be seen that this approximation provides a precision at the level of $1\%$. 

If the photon field is isotropic, the interaction rate can be expressed as
\begin{equation}\label{eq:bb_rate_iso_exp}
\begin{array}{rcl}
{\dot{N}_{\rm iso}}&=& {r_0^2m_{\rm e}^3c^4\kappa \over16\pi\hbar^3E^3}\int\limits_0^\infty\,{db\over{\rm e}^{b/t}-1}\times\\ 
\multicolumn{3}{c}{\int\limits_{0}^{b}\,dx\,\left[\left(1-\frac4x-\frac8{x^2}\right)\log(1+x)+\frac12+\frac8x-\frac1{2(1+x)^2}\right]}\\[10pt]
&\ds=&\ds {2r_0^2m_{\rm e}^3c^4\kappa T^2\over\pi\hbar^3E}F_{\rm n,iso}\left(t\right)\,.
\end{array}
\end{equation}
Here the integral term has properties similar to Equation~(\ref{eq:nani_asym}):
\begin{equation}\label{eq:niso_asym}
\begin{array}{lc}
\ds F_{\rm n,iso}(u)=\left\{{\frac23\zeta(3)\,u=0.801\,u, \atop \frac{\pi^2}{12}\,\log(u)=0.822\,u,}\right.&\ds{ u\ll 1
 \atop  u\gg 1}\,,
\end{array}
\end{equation}
Thus, in the zeroth-order approximation this function can be presented as 
\begin{equation}\label{eq:G0_niso}
G_{\rm n,iso}^{(0)}(u)=0.822\log\left(1+0.97u\right)\,.
\end{equation}
This approximation provides relatively poor precision (at the level of $<30\%$). However, adopting the correction function  given in Equation~(\ref{eq:correction}), one achieves a much higher precision (better than $1\%$, see Figure~\ref{fig:rate}) with the following set of parameters:
\begin{equation}\label{eq:par_iso_rate}
\begin{array}{rclrcl}
\alpha_{\rm n,iso}&=&0.88,& a_{\rm n,iso}&=&0.829,\\
\beta_{\rm n,iso}&=&1.135,& b_{\rm n,iso}&=&1.27\,.
\end{array}
\end{equation}

Combining Equations~(\ref{eq:bb_losses}),~(\ref{eq:bb_rate_ani})~and~(\ref{eq:bb_rate_iso}) one can determine the temperature-dependence of the emitted photon mean energy:
\begin{equation}\label{eq:mean_ani_iso}
\bar{z}_{\rm ani/iso}=\frac1E{\dot{E}_{\rm ani/iso}\over \dot{N}_{\rm ani/iso}}={F_{\rm ani/iso}(u)\over F_{\rm n,ani/iso}(u)}\,
\end{equation}
In the case of a mono-directional photon field the argument of the function in the above equation is $u=2ET(1-\cos\theta)$; in the case of an isotropic photon field $u=4ET$. 

\begin{figure}
\plotone{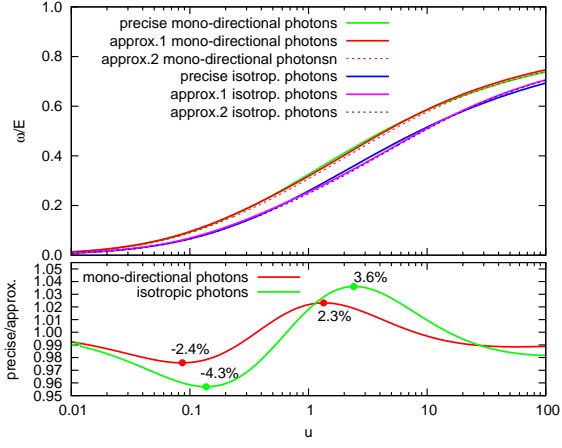}
  \caption{Upper panel: The ratio  of the mean energy of the upscatted photon to the electron energy plotted as a function of $u=t$ (for isotropic photon field) and $u=t_\theta$ (for mono-directional photon field). The approximation 1 corresponds to Equations~(\ref{eq:mean_ani})~and~(\ref{eq:mean_iso}) for the cases of mono-directional and isotropic photon fields, respectively. The approximation 2 corresponds  to Equations~(\ref{eq:one_digit_ani})~and~(\ref{eq:one_digit_iso}) for the cases of mono-directional and isotropic photon fields, respectively. Bottom panel: ratios of numerical calculations to approximations given by Equations~(\ref{eq:mean_ani})~and~(\ref{eq:mean_iso}) for the cases of mono-directional and isotropic photon fields, respectively.}
  \label{fig:mean}
\end{figure}

Obviously, the approximate formulae found for functions $F_{\rm ani}$, $F_{\rm iso}$, $F_{\rm n,ani}$, and $F_{\rm n,iso}$ (see Equations~(\ref{eq:losses_ani_zero}), (\ref{eq:losses_iso_zero}), (\ref{eq:par_iso}), (\ref{eq:G0_nani}), (\ref{eq:par_ani_rate}), (\ref{eq:G0_niso}), (\ref{eq:par_iso_rate})) allow  derivation of high-precision analytical formulae for $\bar{z}$.  However, the  1-parameter freedom in Equations~(\ref{eq:losses_ani_zero}) and (\ref{eq:losses_iso_zero}) allows a significant simplification of the expressions. Namely, in the case of a mono-directional photon field, the mean energy can be approximated as
\begin{equation}\label{eq:mean_ani}
\bar{z}_{\rm ani}={G^{(0)}_{\rm ani}\over G^{(0)}_{\rm n,ani}}\,
\end{equation}
with $c_{\rm z,ani}\simeq4.26$  in $G^{(0)}_{\rm ani}$ that  minimizes the deviation of the approximation function from the precise expression. As seen in Figure~\ref{fig:mean} the error remains below $3\%$. Rounding the coefficients to one non-zero digit (i.e., keeping the precision at the level of $10\%$), one obtains:
\begin{equation}\label{eq:one_digit_ani}
\bar{z}_{\rm ani}={t_\theta\over t_\theta+0.2}{\log(1+t_\theta/2)\over \log(1+2t_\theta)}\,.
\end{equation}

Similarly, for the case of an isotropic photon field, the approximated formula for the fraction of the mean energy,
\begin{equation}\label{eq:mean_iso}
\bar{z}_{\rm iso}={G^{(0)}_{\rm iso}\over G^{(0)}_{\rm n,iso}}\,,
\end{equation}
results in a $5\%$ precision (see Figure~\ref{fig:mean}) for the value of $c_{\rm z,iso}=2.9$ in $G_{\rm iso}^{(0)}$. Rounding the coefficients to one non-zero digit one obtains:
\begin{equation}\label{eq:one_digit_iso}
\bar{z}_{\rm iso}={t\over t+0.3}{\log(1+t/4)\over \log(1+t)}\,.
\end{equation}

The mean photon energy characterizes the typical energy band of upscattered photons in which an electron loses its energy via IC process. Note that in many cases is more demanded the inverse problem, i.e., a reconstruction of the electron energy on the basis of the observed photon energy and the temperature of the target photons. In other words, one needs to solve the transcendental Equation~(\ref{eq:one_digit_iso}) (or Equation~(\ref{eq:one_digit_ani})), to obtain $t$ (or respectively $t_{\theta}$) as a function of $\bar{\omega}$ and $T$. Using Equations~(\ref{eq:one_digit_ani})~and~(\ref{eq:one_digit_iso}) one can derive the following approximate solution:
\begin{equation}\label{eq:one_digit_parent_iso}
t_{\theta/\cdot}={v^{1/2}\left(1+2v^{1/2}\right)\over2}\sqrt{\log(1+v^{1/2})\over \log(1+v^{1/2}/V_0)}\,,
\end{equation}
which allows to obtain the lepton energy with $<10\%$ precision in the entire range of parameters. One should adopt  $V_0=3$ and $v=2\bar{\omega}T(1-\cos\theta)$ for the case of mono-directional photon field; and $V_0=4$ and $v=4\bar{\omega}T$ for isotropic distribution of photons.

\section{Impact of relativistic motion}\label{sec:jet}

The formulae presented in the previous sections correspond to the reference system, where the source of photons is at rest (we refer this system as $K$). Since the particle distribution can be always transformed to this coordinate system, these formulae can, in principle, cover all the required calculations. However, under certain conditions it is more convenient to perform calculations in another coordinate system, $K'$ (the physical quantities measured in this system are marked with prime, e.g. $\omega_0'$). In case if the target photons are mono-directional in the reference frame $K$ (in the relevant region of space), the transformation of the obtained formulae is straightforward. Indeed, in this case the photon distribution function in the 6-dimensional momentum-coordinate phase space ($dN=\rho\,d^3\mathbf{p}d^3\mathbf{r}$) has the following form:
\begin{equation}\label{eq:distr_function_ani}
\rho(\mathbf{p},\mathbf{r})=\delta\left(\mathbf{n}_{\mathbf{p}}-\mathbf{n}_0\right)\frac{c}{p^2}{n_{\rm ph}(\omega_0)}\,,
\end{equation}
{where $\mathbf{n}_{\mathbf{p}}=\mathbf{p}/p$, $\omega_0=cp$, and $\mathbf{n}_0$ is the unit vector corresponding to the direction of photons in the system $K$. Function $n_{\rm ph}$ corresponds to the energy distribution of the target photons, i.e., $dN_{\rm ph}=n_{\rm ph}d\omega_0$, and in the system $K$ is Planckian.}
{The function $\rho$ is a Lorentz invariant \citep[see, e.g.,][]{landau2}, i.e., $\rho(\mathbf{p},\mathbf{r})=\rho'(\mathbf{p}',\mathbf{r}')$, and it can be shown that, in this specific case, the function $n_{\rm ph}$ is an invariant as well:}
\begin{equation}\label{eq:planck_ani}
{n_{\rm ph}'(\omega_0')}={n_{\rm ph}\left(\omega_0\right)}\,. %\over \Gamma\left(1-(v_0/c)\cos\chi\right)
\end{equation}
{Here target photon energies $\omega_0$ and $\omega_0'$ are related via the Lorentz transformation: $\omega_0={\cal D}_*\omega_0'$,  where ${\cal D}_*=\big[\Gamma\left(1-(v_0/c)\cos\chi\right)\big]^{-1}$ is the Doppler factor between the source of blackbody photons (reference system $K$) and the gamma-ray production region (reference system $K'$) moving with relative  velocity $v_0$, that makes an angle   $\chi$ to the photon momentum (the variable $\Gamma=\left(1-(v_0/c)^2\right)^{-1/2}$ denotes the bulk Lorentz factor). Since function $n_{\rm ph}(\omega_0)$ in Equation~(\ref{eq:planck_ani}) is determined by Equation~(\ref{eq:planck}), one can see that in the moving coordinate system $K'$ the Planckian distribution of photons is preserved,  but the temperature of the photon field is corrected for the bulk motion:
\begin{equation}\label{eq:bulk_temerature}
T'={\cal D}_*^{-1}T\,,
\end{equation}
and {an additional dilution factor is applied}:}
\begin{equation}\label{eq:bulk_dillution}
\kappa'={\cal D}_*^2\,.
\end{equation}
Equations~(\ref{eq:bulk_temerature}) and (\ref{eq:bulk_dillution}) allow a generalization of the  formulae obtained in the previous sections for the case of a mono-directional photon field  to a moving system $K'$. As it follows from their derivations, Equations~(\ref{eq:distr_function_ani})~and~(\ref{eq:planck_ani}) {do not account for the relativistic effects related to the transformation of the IC emission from the source frame to the observer frame \citep[for detail see, e.g.,][]{rybicki,jester08}. Also we note that the interaction angle, $\theta$, should also be transformed to the system $K'$. The transformation of the interaction angle can be readily obtained by considering the scalar product of the 4-momenta of electron and the target photon (i.e., a Lorentz invariant quantity). Finally, in the case when the emitting particles are isotropized in the reference frame $K'$, the energy loss rate is described by Equation~(\ref{eq:losses_iso}) with corrections imposed by Equations~(\ref{eq:bulk_temerature})~and~(\ref{eq:bulk_dillution}). }

We leave out of the scope of this paper the transformation of a photon field isotropic in the system $K$ to the moving system $K'$. If looked from the system $K'$, such a photon field is not isotropic field, therefore the basic equation Equation~(\ref{eq:aa_iso}) is not applicable for description the IC scattering process. If the bulk Lorentz factor is large, $\Gamma\gg1$, the photon field in the $K'$  system appears to be nearly mono-directional, with photons moving against the bulk velocity. But the energy distribution of the photons in this case deviates significantly from the Planckian distribution. We note however that in the case of an isotropic photon field, the distribution of electrons can be transformed to the system $K$ and the formulae obtained for the spectrum can be used.

\section{Comparison with other approaches}\label{sec:comp}

In order to simplify calculations, one may try to replace the relatively narrow Planckian distribution by the $\delta$-function, or alternatively use a simplified description of the cross-section, e.g., by the Heaviside step function \citep{petruk09} or simply by a $\delta$-function. For the sake of shortness, in what follows we discuss the $\delta$-functional approximation for a mono-directional target photon field, and the step function approximation by \citet{petruk09} for the scattering off isotropic photon field.

The $\delta$-functional approximation for target photons assumes the following photon field:
\begin{equation}\label{eq:ph_delta}
{dN_{\rm ph}\over d\omega_0}\simeq n_*\delta\left(\omega_0-\omega_*\right)\,.
\end{equation} 
It is easy to be convinced that for  $n_*={2\zeta(3)m_{\rm e}^3c^3\kappa T^3\over\pi^2\hbar^3}$ and $\omega_*=\frac{\pi^4}{30\zeta(3)}\,T\simeq2.7\,T$ one can reproduce correctly both the number and energy densities of the Planckian photon field. However, to a certain extent, the choice of these parameters is arbitrary. 

The substitution of Equation~(\ref{eq:ph_delta}) into Equation~(\ref{eq:bb_ic}) results in the following expression:
\begin{equation}\label{eq:cross_ani_delta}
\begin{array}{l}
\ds\quad\quad{dN_{\rm ani}\over d \omega\,dt}={2r_o^2m_{\rm e}^3c^4\kappa T^2\over \pi\hbar^3E^2}\times\\[10pt]\ds\bigg[{z^2\over2(1-z)}G_{1,\delta}\left(x_0\right)+G_{2,\delta}\left(x_0\right)\bigg]\,,
\end{array}
\end{equation}
where
\begin{equation}\label{eq:f1_delta}
\begin{array}{l}
\ds\quad\quad G_{1,\delta}\left(x_0\right)={n_*\hbar^3\pi^2\over \kappa T^2\omega_*m_{\rm e}^3c^3}\Theta\left(\frac{\omega_*}T-x_0\right)\,,
\end{array}
\end{equation}
and
\begin{equation}\label{eq:f2_delta}
\begin{array}{l}
\ds G_{2,\delta}\left(x_0\right)={n_*\hbar^3\pi^2\over \kappa T^2\omega_*m_{\rm e}^3c^3}\Theta\left(\frac{\omega_*}T-x_0\right)\times\\[10pt]\ds\quad \quad \bigg(1-\frac{2x_0}{\omega_*/T}+\frac{2x_0^2}{\left(\omega_*/T\right)^2}\bigg)\,.
\end{array}
\end{equation}
Here $\Theta\left(x_0\right)$ is the Heaviside step function. 

The comparison of Equations~(\ref{eq:f1_delta})~and~(\ref{eq:f_1}), and of Equations~(\ref{eq:f2_delta})~and~(\ref{eq:f_2}), allows us to estimate the errors introduced by the $\delta$-functional approximation: (1) the lower energy part of the spectrum  ($x_0\ll1$) can be reproduced quite well if the selected parameters satisfy the condition $n_*/(T^2\omega_*)=\kappa m_{\rm e}^3c^3/(6\hbar^3)$; (2) the accuracy declines significantly for the high energy part ($x_0\gtrsim1$).  The $\delta$-function imposes an artificial cutoff at $x_0=\omega_*/T$. Also the accuracy close to the cutoff appears to be quite poor. For example, the accuracy of the term $G_{1,\delta}$ can be estimated as $G_{1,\delta}/G_1^{(0)}$ which gives a factor of $3$ error for $x_0=2$ (assuming that $\omega_*/T>2$, as commonly adopted). 

Similarly, it can be shown that for the cross section averaged over the interaction angle, the $\delta$-functional approximation gives a similar precision. It means that although a practical realization of the $\delta$-functional approximation is characterized by a similar complexity as the approach suggested in this paper, the accuracy provided by the $\delta$-functional approximation is very poor, especially in the Klein-Nishina regime. 

\begin{figure}
\plotone{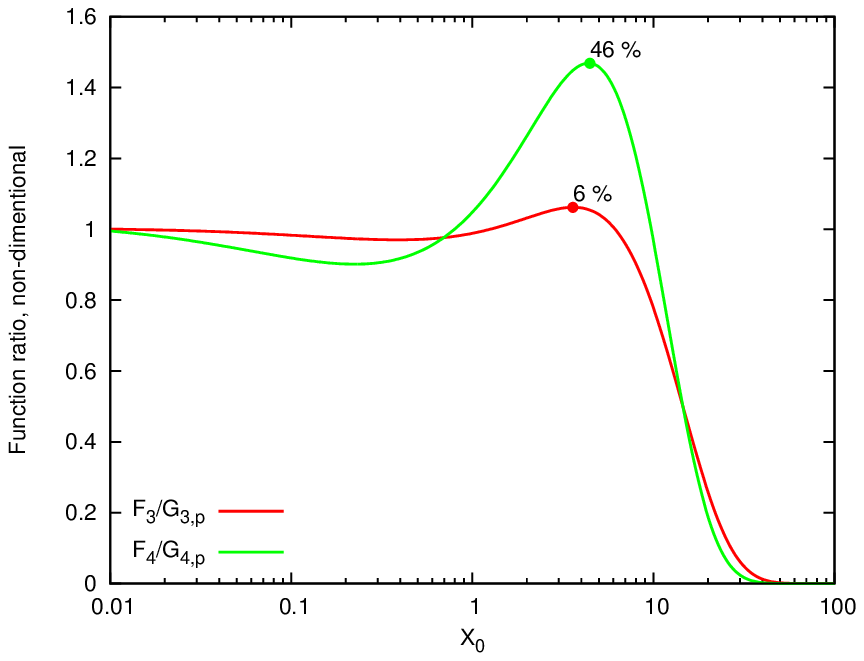}
  \caption{Comparison of the approximation proposed by  \citet{petruk09} to the numerical calculations.}
  \label{fig:petruk}
\end{figure}

A more complicated approach has been suggested by \citet{petruk09} for the case of the cross section averaged over the interaction angle. In this approach the IC cross section was approximated by a step function and integrated over the Planckian photon distribution. This approximation, can be expressed as
\begin{equation}\label{eq:cross_iso_petruk}
\begin{array}{l}
\ds\quad\quad{dN_{\rm iso}\over d \omega\,dt}={2r_o^2m_{\rm e}^3c^4\kappa T^2\over \pi\hbar^3E^2}\times\\[10pt]\ds\bigg[{z^2\over2(1-z)}G_{3,\rm p}\left(x_0\right)+G_{4,\rm p}\left(x_0\right)\bigg]\,,
\end{array}
\end{equation}
where
\begin{equation}\label{eq:f3_petruk}
G_{3,\rm p}\left(x_0\right)=\frac{\pi^2}6{\rm e}^{-\frac23x_0-\frac57x_0^{0.7}}\,,
\end{equation}
and
\begin{equation}\label{eq:f2_petruk}
G_{4,\rm p}\left(x_0\right)=\frac{\pi^2}6{\rm e}^{-\frac23x_0-\frac54x_0^{0.5}}\,.
\end{equation}
In Figure~\ref{fig:petruk} we compare the approximation of \citet{petruk09} with precise numerical calculations, and arrive at a conclusion, similar to the statement of \cite{zdz13}, that for certain parameters this approximation cannot guaranty a precision better than $\sim50\%$ \citep[we note that][did not approximate the cross section in the region of the exponential tail, i.e. for $x_0\gg1$]{petruk09}.

\begin{figure}
\plotone{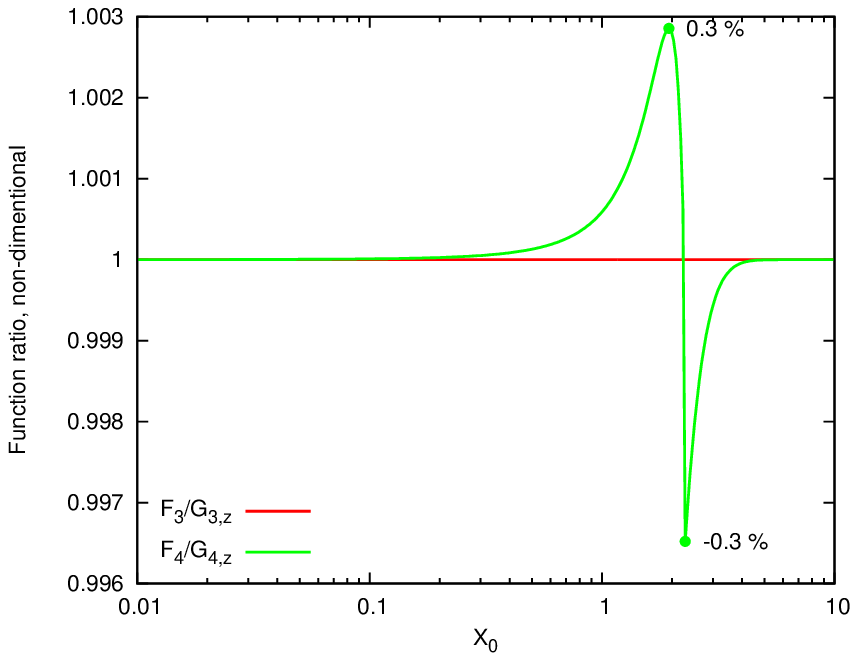}
  \caption{Comparison of the approximation proposed by  \citet{zdz13} to the numerical calculations.}
  \label{fig:zdziarski}
\end{figure}

Finally, \citet{zdz13} suggested an analytical method to describe
Equations~(\ref{eq:aa_ani})~and~(\ref{eq:aa_iso}) based on the truncation
of series, which describe the functions $f_{-1}$ and $f_{\rm
  ln}$. { In Figure~\ref{fig:zdziarski} we compare precise numerical
  calculations to the approximated values, $G_{3,z}$ and $G_{4,z}$,
  that were obtained by substitution of Equations~(7), (8), (14), and
  (28) from \citet{zdz13} (the value of $N=3$, as suggested by the
  authors, was adopted) to
  Equations~(\ref{eq:f_3})~and~(\ref{eq:f_4}). Since \citet{zdz13}
  obtained analytical expressions for the functions $f_{+1}$ and
  $f_{0}$, the function $G_{3,z}$ is strictly equal to $F_3$, which can be seen
  in Figure~\ref{fig:zdziarski}. The accuracy provided by the function
  $G_{4,z}$ is very high, at the level of $0.3\%$\footnote{ This accuracy is worse by
    approximately a factor of 10 than the accuracy of
    $<2.4\times10^{-4}$ achieved for the functions $f_{-1}$ and
    $f_{\rm ln}$ \citep{zdz13}.  This discrepancy is explained by the
    fact that these functions enter into the expression for the cross
    section with different signs, and the subtraction of these
    functions results in an overall error significantly exceeding the
    accuracy of the individual terms.} (see
  Figure~\ref{fig:zdziarski}). }

{ The approach by \citet{zdz13} can provide an
  arbitrary precision (simply by increasing of the number the preserved
  terms in the series), however, in our view, it also owns a certain
  shortcoming. Namely, this approach implies the usage of
  special functions (dilogarithm and exponential integral), which may
  harden the practical usage.}

Another important difference of our approach is that while in other
studies one provides an approximate description for the cross section, we
suggest an approach for a common description of all the relevant processes
of IC scattering on the black-body photons: scattering rates,
energy losses, cross sections and mean photon energy. Also, all the
approximations use the same type of correction function,
Equation~(\ref{eq:correction}).

\section{Summary}
In this paper we suggest simple analytical presentations for calculations of
different characteristics (differential spectra, interaction rates,
and energy losses) of the IC scattering of relativistic electrons in
the radiation field which is described by Planckian distribution.
Two different types of angular distribution, namely mono-directional
and isotropic distributions of the target radiation field have been
considered.

The obtained parameterizations  are characterized by a high precision, of an order of $1\%$, and cover the
entire parameter space allowing an accurate description of the IC
scattering in the Thomson and Klein-Nishina limits, as well as in the
transition region. The derived formulae preserve the precise
asymptotic behavior and have similar structures, which simplifies their
practical usage (see Table \ref{table:parameters}). 

The main objective of the obtained approximate analytical
presentations is the fast, but convenient and accurate calculations of
characteristics of the upscattered IC emission in radiation fields
described by Planckian distribution. At the same time, the simple
forms of these parameterizations allow derivation of some useful
relations.  In particular, we propose simple formulae which, for the
given temperature of target photons, relate the mean energy of the
electron and up-scattered photon.


\begin{deluxetable}{c c c c c c c c c c c}  
\tabletypesize{\scriptsize}
\rotate
\tablewidth{0pt}
\tablecaption{Derived parameterizations\label{table:parameters}} 
\tablehead{
 Function & Used in Eqs. & \multicolumn{2}{c}{\,\,\,Approximated description} & \multicolumn{5}{c}{Fit parameters values} &   precision & Figure\\
  \hline
   & &function & variable & $a$ & $\alpha$ & $b$ & $\beta$ & $c$ &   &\\
}
\startdata
  \multicolumn{11}{c}{Representation of Spectra}\\
  \hline
\\
  $F_{1}$ & Eq.~(\ref{eq:cross_ani})&$G^{(0)}_{1}=\left(\frac{\pi^2}6+x_0\right){\rm     e}^{-x_0}$& ${z\over1-z}{1\over2ET(1-\cos\theta)}$ &  --- & --- & --- & --- & --- & $\sim10\%$ &   Figs.~\ref{fig:f_12},\ref{fig:f_1ratio}\\
  $F_{1}$ & Eq.~(\ref{eq:cross_ani})&$G^{(0)}_{1}\left(x_0\right)\times g(x_0)$& ${z\over1-z}{1\over2ET(1-\cos\theta)}$  & $0.153$ & $0.857$ & $0.254$ & $1.84$& --- & $<1\%$   & Figs.~\ref{fig:f_12},\ref{fig:f_1ratio}\\
  $F_{2}$ & Eq.~(\ref{eq:cross_ani})&$G^{(0)}_{2}=\left(\frac{\pi^2}6+x_0\right){\rm     e}^{-x_0}$&${z\over1-z}{1\over2ET(1-\cos\theta)}$ & --- & --- & ---   & --- & --- & $\sim50\%$ & Figs.~\ref{fig:f_12},\ref{fig:f_2ratio}\\
  $F_{2}$ & Eq.~(\ref{eq:cross_ani})&$G^{(0)}_{2}\left(x_0\right)\times g(x_0)$ &${z\over1-z}{1\over2ET(1-\cos\theta)}$ & $1.33$ &   $0.691$ & $0.534$ & $1.668$& --- & $<1\%$ & Figs.~\ref{fig:f_12},\ref{fig:f_2ratio}\\
  $F_{3}$ & Eq.~(\ref{eq:cross_iso})&$G^{(0)}_{3}=\frac{\pi^2}6\frac{1+c\,x_0}{1+\frac{\pi^2c}6x_0}{\rm     e}^{-x_0}$&${z\over1-z}{1\over4ET}$ & --- & --- & --- & --- & $2.73$ & $\sim3\%$ &   Figs.~\ref{fig:f_34},\ref{fig:f_3ratio}\\
  $F_{3}$ & Eq.~(\ref{eq:cross_iso})&$G^{(0)}_{3}\left(x_0\right)\times g(x_0)$&${z\over1-z}{1\over4ET}$  & $0.443$ & $0.606$ & $0.54$ & $1.481$& $0.319$ & $<1\%$   & Figs.~\ref{fig:f_34},\ref{fig:f_3ratio}\\
  $F_{4}$ & Eq.~(\ref{eq:cross_iso})&$G^{(0)}_{4}=\frac{\pi^2}6\frac{1+c\,x_0}{1+\frac{\pi^2c}6x_0}{\rm     e}^{-x_0}$&${z\over1-z}{1\over4ET}$  & --- & --- & ---   & --- & $47.1$ & $\sim30\%$ & Figs.~\ref{fig:f_34},\ref{fig:f_4ratio}\\
  $F_{4}$ & Eq.~(\ref{eq:cross_iso})&$G^{(0)}_{4}\left(x_0\right)\times g(x_0)$&${z\over1-z}{1\over4ET}$  & $0.726$ &   $0.461$ & $0.382$ & $1.457$& $6.62$ & $<1\%$ & Figs.~\ref{fig:f_34},\ref{fig:f_4ratio}\\
  \hline                             
\\
  \multicolumn{11}{c}{Energy Losses}\\
  \hline
\\
  $F_{\rm ani}$ & Eq.~(\ref{eq:losses_ani}, \ref{eq:mean_ani})&$G^{(0)}_{\rm ani}={cu\log(1+2.16u/c)\over1+cu/0.822}$& $2ET(1-\cos\theta)$ & --- & --- & --- & ---& $6.13$ & $\sim1\%$ &   Fig.~\ref{fig:losses}\\
  $F_{\rm iso}$ & Eq.~(\ref{eq:losses_iso}, \ref{eq:mean_iso})&$G^{(0)}_{\rm iso}={cu\log(1+0.722u/c)\over1+cu/0.822}$&$4ET$ & --- & --- & --- & ---& $4.62$ & $\sim5\%$ &   Fig.~\ref{fig:losses}\\
  $F_{\rm iso}$ & Eq.~(\ref{eq:losses_iso})&$G^{(0)}_{\rm iso}\left(u\right)\times g(u)$&$4ET$ & $-0.362$ & $0.682$ & $0.826$ & $1.281$& $5.68$ & $\sim1\%$ &   Fig.~\ref{fig:losses}\\
  \hline                             
\\
  \multicolumn{11}{c}{Interaction Rate}\\
  \hline
\\
  $F_{\rm n,ani}$ & Eq.~(\ref{eq:bb_rate_ani_exp}, \ref{eq:mean_ani})&$G^{(0)}_{\rm n,ani}=0.822\log\left(1+1.949u\right)$ &$2ET(1-\cos\theta)$ & --- & --- & --- & --- & --- & $\sim25\%$ &   Fig.~\ref{fig:rate}\\
  $F_{\rm n,ani}$ & Eq.~(\ref{eq:bb_rate_ani_exp})&$G^{(0)}_{\rm n,ani}\left(u\right)\times g(u)$ & $2ET(1-\cos\theta)$ & $1.05$ & $0.885$ & $2.46$ & $1.213$& --- & $\sim1\%$ &   Fig.~\ref{fig:rate}\\
  $F_{\rm n,iso}$ & Eq.~(\ref{eq:bb_rate_iso_exp}, \ref{eq:mean_iso})&$G^{(0)}_{\rm n,iso}=0.822\log\left(1+0.97u\right)$& $4ET$ & --- & --- & --- & --- & --- & $\sim30\%$ &   Fig.~\ref{fig:rate}\\
  $F_{\rm n,iso}$ & Eq.~(\ref{eq:bb_rate_iso_exp})&$G^{(0)}_{\rm n,iso}\left(u\right) \times g(u)$& $4ET$& $0.829$ & $0.88$ & $1.27$ & $1.135$& --- & $\sim1\%$ &   Fig.~\ref{fig:rate}\\
  \hline                             
\\
  \multicolumn{11}{c}{Mean Energy of Emitted Photons }\\
  \hline
\\
  $\bar{z}_{\rm ani}$ & Eq.~(\ref{eq:mean_ani})&$G^{(0)}_{\rm ani}/G^{(0)}_{\rm n,ani}$ &$2ET(1-\cos\theta)$ & --- & --- & --- & ---& $4.26$ & $\sim3\%$ &   Fig.~\ref{fig:mean}\\
  $\bar{z}_{\rm ani}$ & Eq.~(\ref{eq:one_digit_ani})&${u\over u+0.2}{\log(1+u/2)\over \log(1+2u)}$ &$2ET(1-\cos\theta)$ & --- & --- & --- & ---& --- & $\sim8\%$ &   Fig.~\ref{fig:mean}\\
  $\bar{z}_{\rm iso}$ & Eq.~(\ref{eq:mean_iso})&$G^{(0)}_{\rm iso}/G^{(0)}_{\rm n,iso}$ &$4ET$  & --- & --- & --- & ---& $2.9$ & $\sim5\%$ &   Fig.~\ref{fig:mean}\\
  $\bar{z}_{\rm iso}$ & Eq.~(\ref{eq:one_digit_iso})&${u\over u+0.3}{\log(1+u/4)\over \log(1+u)}$ &$4ET$  & --- & --- & --- & ---& --- & $\sim8\%$ &   Fig.~\ref{fig:mean}\\
\hline
\\
  \multicolumn{11}{c}{Energy of Emitting Particle}\\
  \hline
\\
  $t_{\theta}$ & Eq.~(\ref{eq:one_digit_parent_iso})&${v^{1/2}\left(1+2v^{1/2}\right)\over2}\sqrt{\log(1+v^{1/2})\over \log(1+v^{1/2}/3)}$ &$2\bar{\omega}T(1-\cos\theta)$& --- & --- & --- & ---& --- & $\sim10\%$ &   ---\\
  $t$ & Eq.~(\ref{eq:one_digit_parent_iso})&${v^{1/2}\left(1+2v^{1/2}\right)\over2}\sqrt{\log(1+v^{1/2})\over \log(1+v^{1/2}/4)}$ &$4\bar{\omega}T$ & --- & --- & --- & ---& --- & $\sim10\%$ &   ---\\
\hline
\enddata
\tablecomments{A typical parametrization consists of a zeroth-order approximation function, $G^{(0)}_{\cdot\cdot\cdot}$, multiplied by the correction factor $g$. The zeroth-order approximation  depends on the variable, which is listed in the fourth column of the table, and in some cases on the parameter $c$. The correction factor is given by Equation~(\ref{eq:correction}), and depends on the same variable as the zeroth-order approximation, and four parameters ($a$, $\alpha$, $b$ and $\beta$).}
\end{deluxetable}

\end{document}